\documentstyle[aps,epsf,epsfig,psfig,float,latexsym]{revtex}

\newcommand{\beq}{\begin{equation}}
\newcommand{\eeq}{\end{equation}}
\newcommand{\beqa}{\begin{eqnarray}}
\newcommand{\eeqa}{\end{eqnarray}}

\newcommand{\no}{\nonumber}

\newcommand{\vs}{\vspace{-0.0cm}}
\newcommand{\fet}[1]{\mbox{\boldmath $#1$}}

\begin{document}

\title{
\begin{flushright}
{\protect \tiny {FZJ-IKP(TH)-2001-09}}\\
{\normalsize{}}
\end{flushright}
\vspace{1 cm}
Resonance saturation for four--nucleon operators
}
\vspace{0.5 cm}
\author{
Evgeny~Epelbaum$^{a}$\footnote{email: evgeni.epelbaum@tp2.ruhr-uni-bochum.de},
Ulf-G. Mei{\ss}ner$^a$\footnote{email: u.meissner@fz-juelich.de},
Walter~Gl\"ockle$^b$\footnote{email: walter.gloeckle@tp2.ruhr-uni-bochum.de},
Charlotte Elster$^{a,c}$\footnote{email: c.elster@fz-juelich.de}
\\[0.3em]
{\small {$^a${\it Forschungszentrum J{\" u}lich, Institut f{\" u}r Kernphysik
(Th), D-52425 J{\" u}lich, Germany}}\\
$^b${\it Ruhr-Universit\"at Bochum, Institut f{\"u}r
  Theoretische Physik II, D-44870 Bochum, Germany}}\\
$^c${\it Instiute for Nuclear and Particle Physics, Ohio
      University, Athens, OH 45701, USA}}
%}

\maketitle

\vspace{0.5cm}

\begin{abstract}
\noindent In the modern description of nuclear forces based on chiral effective
field theory, four--nucleon operators with unknown coupling constants
appear. These couplings can be fixed by a fit to the low partial
waves of neutron--proton scattering. We show that the so determined
numerical values can be understood on the basis of phenomenological
one--boson--exchange models. We also extract these values from various modern
high accuracy nucleon--nucleon potentials and demonstrate their consistency and
remarkable agreement with the values in the chiral effective field theory
approach.
This paves the way for estimating the
low--energy constants of operators with more nucleon fields and/or
external probes.

\vspace{0.2cm}


\end{abstract}

\vspace{1cm}

%\vspace{1cm}

\section{Introduction}\label{sec:intro}
\noindent
Effective chiral Lagrangians can be used to investigate
the dynamics of pion, pion--nucleon as well as nucleon--nucleon interactions.
In all cases, one has to consider two distinct contributions, namely tree
and loop diagrams, which are organized according to the underlying
power counting \cite{wein,GL85}. To a given order, one has to consider all
local operators
constructed from pions, nucleon fields and external sources in harmony with
chiral symmetry, Lorentz invariance and the pertinent discrete symmetries.
Beyond (or even at) leading order in the chiral expansion, these operators are
accompanied by unknown coupling constants, also
called low--energy constants (LECs). In principle, these LECs are calculable
from QCD but in practice need to be fixed by a fit to some data or using
some model~\footnote{These LECs can also be calculated in lattice gauge theory.
For a first attempt in the Goldstone boson sector using the strong coupling
expansion see ref.~\cite{MR} while the most recent quenched calculation
for these LECs is given in~\cite{alphaC}.}. While in certain cases
sufficient data exist allowing one to pin down the LECs, often some
reliable estimate for these constants beyond naive dimensional analysis
is needed. In the meson sector, the ten LECs of the chiral Lagrangian at
next--to--leading order (NLO) have been determined \cite{GL85} and their values
can be
understood in terms of masses and coupling constants of the lowest meson
resonances of vector, axial--vector, scalar and pseudoscalar character,
may be with the exception of the scalar sector with vacuum quantum
numbers \cite{EGPdR,DRV}.
This is called resonance saturation, it has been used e.g. to estimate
LECs at next--to--next--to--leading order (NNLO) (see e.g. \cite{2loop}) or
for the extended chiral Lagrangian including virtual photons as dynamical
degrees of freedom \cite{BU} \footnote{For a critical discussion of resonance
saturation concerning these LECs, see \cite{bashir}. Note also
that the status of resonance saturation  for the
non--leptonic weak LECs is less clear \cite{KMW}.}. A similar systematic
analysis exists
for the finite dimension two couplings of the pion--nucleon effective
Lagrangian
\cite{BKMasp},
where it was demonstrated the LECs are saturated in terms of baryon resonance
excitation in the s-- and u--channel and t--channel meson resonances. Much less
is known about dimension three and four couplings, but for certain processes
resonance saturation has been shown to work quite well,  e.g. in neutral
pion photoproduction off protons \cite{BKMphoto}.
A somewhat different scheme (including also meson--resonance loops) was
introduced in the study of the baryon octet masses in
ref.~\cite{BM}. The situation is very different
concerning few--nucleon systems, where a new type of operators with $2A$
nucleon
fields appears (for reactions involving $A\ge 2$ nucleons). Only recently, a
complete
and precise determination of the four S--wave and five P--wave (LO and NLO)
LECs in neutron--proton scattering has become available \cite{EGMII},
thus it is timely to ask the question
whether the numerical values of these four--nucleon coupling constants can
be understood from some kind of resonance saturation.\footnote{Note that in the
previous similar work~\cite{bira} global fits with 26 free parameters
where performed, which presumably do not allow to pin down the LECs in a unique
way.
For more details and further discussion on various differences between our
formalism
and the one of the ref.~\cite{bira} see \cite{EGMI,EGMII}.}
This will be the topic
of the present paper. As it will turn out, existing one--boson--exchange
(or more phenomenologically constructed) models of the nucleon--nucleon (NN)
force allow to explain the LECs in terms
of resonance parameters.  This paves the way for estimating the
low--energy constants of operators with more nucleon fields and/or
external probes. Before elaborating on the details, we mention that in
studies of pion production in proton--proton collisions or charge symmetry
breaking in the NN interaction, ideas of resonance saturation have already
been used \cite{vKMR,vKFG}. Also, Friar~\cite{F} has discussed aspects
of integrating out heavy meson fields to generate local four--nucleon
operators with given LECs but did not attempt a detailed comparison with
existing models of the nuclear forces as done here.

\medskip
\noindent The plan of the article
is as follows. In Section~\ref{sec:eft} we discuss the effective chiral
Lagrangian for nucleon--nucleon interactions, in particular the four--nucleon
terms and their corresponding couplings constants. We then summarize how these
LECs are determined at NLO and give a novel prescription to calcuate the NNLO
chiral effective field theory (EFT) potential. In Section~\ref{sec:pot} we show
how to calculate these LECs from existing boson--exchange or  phenomenological
potentials and compare the resulting values with the ones obtained in EFT.
Section~\ref{sec:nat} is devoted to the study of naturalness of these
coupling constants and the implications of Wigner's spin--isospin symmetry.
Our conclusions are summarized in Section~\ref{sec:summ}. Some technicalities
are relegated to the appendices.

\section{Chiral Effective Field Theory}
\label{sec:eft}
\subsection{Effective Lagrangian and definition of LECs}
\label{sec:fund}
\noindent
To be specific, we  briefly discuss the approach to
chiral Lagrangians for few--nucleon systems proposed by Weinberg.
One starts from an effective chiral Lagrangian  of
pions and nucleons, including in particular local
four--nucleon interactions which describe the short range part of
the nuclear force, symbolically
\beq\label{L}
{\mathcal L}_{\rm eff}  = {\mathcal L}_{\pi\pi} + {\mathcal L}_{\pi N} +
 {\mathcal L}_{NN}~,
\eeq
where each of the terms admits an expansion in small momenta and quark (meson)
masses. To a given order, one has to include all terms consistent with chiral
symmetry, parity, charge conjugation and so on.
The last term in eq.(\ref{L}) contains the four--, six--, $\ldots$ nucleon
terms of
interest here. From the effective Lagrangian, one
derives the two--nucleon potential. This potential is based on (a modified)
Weinberg
counting~\cite{EGMI}, more precisely, one organizes the unitarily transformed
infrared non--singular diagrams according to their power (chiral dimension)
in  small momenta and pion masses (for a detailed discussion, see
ref.~\cite{EGMI}).
To leading order (LO), this potential is the sum of one--pion exchange (OPE)
(with  point-like coupling)  and of
two four--nucleon contact interactions without derivatives. The
low--energy constants  accompanying these terms have to be
determined by a fit to some data, like e.g. the two S-wave phase shifts in the
low--energy region (for $np$). At
next--to--leading order (NLO), one has corrections to the OPE, the leading
order two--pion exchange graphs and seven dimension
two four--nucleon terms with unknown LECs (for the $np$ system).
Finally, at NNLO, one has further renormalizations of the one-- and
corrections to the
two--pion exchange graphs including dimension two pion--nucleon
operators. The corresponding LECs can be determined from the chiral
perturbation theory (CHPT) analysis of pion--nucleon scattering.
The existence of shallow nuclear bound states (and large scattering lengths)
forces one to perform an additional nonperturbative resummation. This is done
here by obtaining the bound and scattering states from the solution of the
Lippmann--Schwinger equation. The potential has to be understood as
regularized, and the regularization is dictated by the EFT approach employed
here, i.e.
\beq
V( {p}, {p}'\,) \to f_R ( {p} ) \, V( {p}, {p}') \, f_R ({p}' )~,
\eeq
where $f_R (p)$ is a regulator function chosen in harmony with the
underlying symmetries. Within a certain range of cut--off values, the physics
should be
independent of its precise form and  value~\cite{lepage}. That this is indeed
the
case has been demonstrated in~\cite{EGMII}. The central object of the
study presented here are the LECs related to the four--nucleon
operators. In a spectroscopic notation these are called $C_{1S0}$,
 $\tilde{C}_{1S0}$,
$C_{3S1}$,  $\tilde{C}_{3S1}$, $C_{3D1-3S1} := C_{\epsilon 1}$, $C_{1P1}$,
$C_{3P0}$,
$C_{3P1}$, and $C_{3P2}$. In the following, we will collectively
denote these as $C_i$ and $\tilde{C}_i$, respectively. The two LECs
$\tilde{C}_i$
stem from the two momentum independent four--nucleon operators while the seven
$C_i$ are related to two--derivative operators as they appear in the
effective Lagrangian (we have adopted the notation to the two--nucleon
potential
given in~\cite{EEdiss}),
\beqa\label{Leff}
{\cal L}_{NN} &=& {\cal L}_{NN}^{(2)} + {\cal L}_{NN}^{(4)} + \ldots,
\no \\
{\cal L}_{NN}^{(2)} &=& -\frac{1}{2}\, C_S \, (N^\dagger N) \, (N^\dagger
N)
-\frac{1}{2}\, C_T \, (N^\dagger \sigma_i \,N) (N^\dagger
\sigma_i \,N) ~, \no \\
{\cal L}_{NN}^{(4)}
&=& {} - \frac{1}{2} {C}_1  \left[  ( N^\dagger  \partial_i N )^2  +
(( \partial_i N^\dagger ) N )^2  \right]
-  \left( {C}_1 - \frac{1}{4} {C}_2 \right)
( N^\dagger \partial_i N ) (( \partial_i N^\dagger ) N ) \nonumber \\
& & {} + \frac{1}{8} {C}_2  ( N^\dagger N ) \left[ N^\dagger
\partial_i^2 N + \partial_i^2
N^\dagger N \right]  \nonumber \\
& & {} - \frac{i}{8} {C}_5  \epsilon_{ijk} \bigg\{
\left[ ( N^\dagger \partial_i N ) (( \partial_j N^\dagger )
 \sigma_k N ) + (( \partial_i N^\dagger ) N )  ( N^\dagger
\sigma_j
\partial_k N ) \right]  \nonumber \\
& & {} -  ( N^\dagger N ) (( \partial_i N^\dagger ) \sigma_j
\partial_k N )  +   ( N^\dagger \sigma_i N )
(( \partial_j N^\dagger )
 \partial_k N ) \bigg\} \no \\
& & {} + \frac{1}{4} \left( \left( C_6 + \frac{1}{4} C_7
  \right)    \left( \delta_{i k}
\delta_{j l}
+ \delta_{i l} \delta_{k j} \right) +
\left( 2 C_3 + \frac{1}{2} C_4 \right) \delta_{i j} \delta_{k l} \right)
\nonumber \\
& & {} \quad \quad \quad \times \left[  ( (\partial_i \partial_j N^\dagger
)
\sigma_k  N )
+  ( N^\dagger \sigma_k \partial_i \partial_j N ) \right] ( N^\dagger
\sigma_l N )
\nonumber \\
& & {} -  \frac{1}{2} \left( {C}_{6} \left( \delta_{i k} \delta_{jl}
+ \delta_{i l} \delta_{k j} \right) + C_4
 \delta_{i j} \delta_{k l} \right)
( N^\dagger \sigma_k \partial_i N ) (( \partial_j N^\dagger ) \sigma_l N
 )  \nonumber \\
& & {} - \frac{1}{8} \left( \frac{1}{2}  C_7
\left( \delta_{i k} \delta_{j l}
+ \delta_{i l} \delta_{k j}
\right) - ( 4 C_3 - 3 C_4 )
 \delta_{i j} \delta_{k l} \right)
 \left[  ( \partial_i N^\dagger \sigma_k
\partial_j N )
 + ( \partial_j N^\dagger
\sigma_k \partial_i N ) \right] ( N^\dagger \sigma_l N )\,,
\eeqa
where $N$ denotes the (non--relativistic) nucleon fields, $N =
(p,n)^T$, and $\sigma_l$ $(l=1,2,3)$ are the Pauli spin matrices,
and the summation convention for repeated indices is understood.
Since we are not considering external sources here, we only have
partial derivatives acting on the nucleon fields.
To arrive at  this expression for the most general effective Lagrangian
with four nucleon field operators, we have made use of partial integration,
Fierz transformation and the equation of motion for the nucleons. We also
require
reparametrization invariance  \cite{Luke} of the Lagrangian, which allows us
to further reduce the number of independent terms as compared to \cite{bira}.
The complete derivation of eq.~(\ref{Leff}) within the heavy baryon formalism
is
presented in \cite{EEdiss}. Note that the effective Lagrangian (\ref{Leff})
corresponds
to the rest--frame system of the nucleon with the velocity operator given by
$v_\mu = (1, 0, 0, 0)$. The resulting NN contact
 potential reads (in  the centre--of--mass (c.m.)~system):
\begin{eqnarray}
\label{2n6}
V_{{\rm cont}} &=& C_S + C_T ( \vec \sigma_1 \cdot \vec \sigma_2 ) + C_1 {\vec
q \,}^2
+ C_2 {\vec k \,}^2 + ( C_3 {\vec q \, }^2 + C_4 {\vec k \,}^2 )
( \vec \sigma_1 \cdot \vec \sigma_2 )
+ i C_5 \frac{ \vec \sigma_1 + \vec \sigma_2}{2}
\cdot ( \vec q \times \vec k ) \nonumber \\
&& {} + C_6 ( \vec q \cdot \vec \sigma_1 )
( \vec q \cdot \vec \sigma_2 ) + C_7 ( \vec k \cdot \vec \sigma_1 )
( \vec k \cdot \vec \sigma_2 ) \,,
\end{eqnarray}
where $\vec {q} = \vec {p}\, ' - \vec {p} $ and $\vec k = (\vec p + \vec p \,
')/2$ are the
transfered and the averaged momentum, respectively, and $\vec{p}$ ($\vec p \,
'$)  corresponds to
the initial (final) momentum of the nucleons in the c.m.~system.
Closer inspection of  eq.~(\ref{2n6}) might lead
to the question
why no operators containing the isospin matrices $\tau_i$ (where $i=1,2$ labels
the
nucleons) appear? For example, $\rho$--meson exchange will naturally lead to
a contribution $\sim \tau_1 \cdot \tau_2$. In principle, at NLO, one can write
down
18 operators in the effective potential and not just 9 as appear here.
What seems to be completely missing are
the nine operators involving products of isospin matrices. However, we remind
the
reader that only 9 of these 18 operators are independent.
The terms in the Lagrangian related to
the other 9 can be eliminated
{}using Fierz transformations~\cite{wein,EGMI}. Equivalently, one
can perform an antisymmetrization of the two--nucleon potential to eliminate
redundant
terms as used in \cite{KBW,EGMII,EEdiss}. Clearly, the set of
operators we choose to work with is one
but not the unique possibility.

\medskip\noindent
As stated before, there are two/seven LECs
related to operators with zero/two derivatives. These constants
can be most easily determined by a fit to the S-- and P--wave phase
shifts and the $^3S_1-^3D_1$ mixing parameter at low energies, which
leads naturally to certain linear combinations, i.e. the already enumerated
spectroscopic LECs. The precise relation of the  LECs appearing in the
effective Lagrangian to the spectroscopic ones
is taken from ref.\cite{EGMII} (correcting some
typographical errors  in that reference),
\beqa\label{VC}
\tilde{C}_{1S0}  &=& {4\pi} \, (C_S-3C_T)~, \no\\
C_{1S0} &=&  \pi \, ( 4C_1 + C_2 -12C_3
-3C_4 -4C_6 -C_7)~,\no\\
\tilde{C}_{3S1} &=& {4\pi} \, (C_S+C_T)~, \no \\
C_{3S1} &=& \frac{\pi}{3} \, ( 12C_1 + 3C_2 +12C_3
+3C_4 +4C_6 +C_7) ~,\no\\
C_{1P1} &=& \frac{2\pi}{3} \, ( -4C_1 + C_2 +12C_3
-3C_4 +4C_6 -C_7) ~, \no \\
C_{3P1} &=& \frac{2\pi}{3} \, ( -4C_1 + C_2 - 4C_3
+C_4 + 2C_5 -8C_6 + 2C_7)~\no\\
C_{3P_2} &=& \frac{2\pi}{3} \, ( -4C_1 + C_2 - 4C_3
+C_4 - 2C_5 )~, \no \\
C_{3P0} &=& \frac{2\pi}{3} \, ( -4C_1 + C_2 - 4C_3
+C_4 + 4C_5 +12C_6 - 3C_7)~, \no \\
C_{3D1 - 3S1} &=& C_{\epsilon 1} = \frac{2\sqrt{2}\pi}{3} \, ( 4C_6 + C_7)~.
\label{VCend}
\eeqa

\subsection{LECs at next--to--leading order}\label{sec:LNLO}
\noindent
Let us now discuss the determination of NLO LECs of
the chiral EFT potential. In contrast to what was done
in~\cite{EGMII}, we also include the leading charge dependence effect,
which is the charged to neutral pion mass difference,
$\Delta M_\pi = M_{\pi^\pm} - M_{\pi^0}$, in the OPE potential (for a
systematic study of such effects, see~\cite{WME}.) Fitting the low
neutron--proton (np) partial waves (S,P and the triplet S-D mixing)
for center--of--mass energies below 50-100~MeV,
one obtains the numerical values of the LECs for the given regulator
and cut--off value. We work here with an exponential regulator,
\beq
f_R (p) = \exp \left( -p^4 / \Lambda^4 \right) ~,
\eeq
where the momentum cut--off $\Lambda$ is varied between
500 and 600~MeV (a more detailed discussion of various regulator
functions is given in \cite{EGMII}). Therefore, we obtain
a range of values for each LEC in the given partial waves. For a direct
comparison with one--boson--exchange models, we need to further add
the TPE contribution which stems from the box, triangle and
football diagrams. This is done by expanding the contributions of
these graphs in terms of local operators with increasing powers of
derivatives and projecting onto the appropriate partial waves (this
method is described in more detail below).
The so obtained numerical contributions in each partial wave are listed in
table~\ref{tab1} in the column TPE(NLO) (for explicit analytical
expressions, see appendix~\ref{app:TPE}). Obviously, these numbers
are cut--off independent. In contrast, the OPE is retained because
all potentials we will compare to include it as well. We note that some
of these potentials contain a pion--nucleon form factor, but since it
only depends on the momentum transfer squared and appears quadratically,
it does not influence any four--nucleon operator with zero or two derivatives.
With this in mind, we present in table~\ref{tab1} the resulting values of the
LECs
for the cut--off varying from 500 to 600~MeV. This is the optimal range found
in the study of few--nucleon system~\cite{EGMII,EGMIII} as well as
proton--proton scattering~\cite{WME} in the framework used here.
Note that in principle we could take smaller values for the cut--off. In such a
case
one would get a slightly less precise description of the data. On the contrary,
one could not substantially increase the cut--off values if no unphysical
deeply
bound states are allowed~\cite{EGMII,EGMcycle}.\footnote{Although, as will be
stressed in section~\ref{sec:LNNLO},
such spurious bound states would not affect low--energy NN observables,
the direct comparison with the realistic NN potentials would not be possible.}

\subsection{Phase equivalent potentials and LECs at
  next--to--next--to--leading order}\label{sec:LNNLO}

\begin{figure}[tb]
\centerline{\epsfig{file=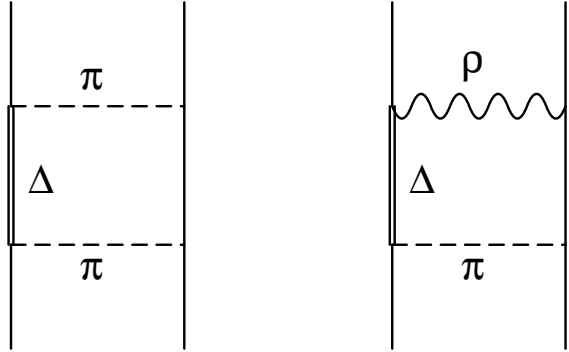,width=3in}}

\vspace{1.3cm}

\caption{\protect \small
Classes of diagrams which cancel to large extent. One representative
TPE graph and one $\pi \rho$ graph are shown. Solid, double, dashed
and wiggly lines represent nucleons, deltas, pions and $\rho$--mesons,
respectively. }\label{fig:1}
\end{figure}

\noindent
We now turn to the determination of the LECs based
on the NNLO potential. Here, we perform a modification as compared to
the work presented in ref.~\cite{EGMII}. At that order, the
pion--nucleon LECs $c_{1,3,4}$ appear, which have been taken from
the CHPT analysis of $\pi N$ scattering in the interior of the Mandelstam
triangle~\cite{Paul}. As already shown in ref.~\cite{BKMasp},
these values can be understood
in terms of baryon and meson resonance excitations, with a particularly
strong contribution from the $\Delta (1232)$ resonance. While the
natural size for these LECs is 1~GeV$^{-1}$, typical values found
for $c_3$ and $c_4$ from $\pi N$ scattering data
are $c_3 = (-4.70\pm 1.16)$~GeV$^{-1}$
and $c_4 = (3.40\pm 0.04)$~GeV$^{-1}$, respectively. The resulting
TPE with insertion of these operators improves the fit but leads to
a very strongly attractive central potential, as witnessed
by the appearance of deeply bound states e.g. in the deuteron channel.
These states do, however, not influence the low--energy physics in the
two--nucleon system. However, the resulting potential is clearly not
phase--equivalent to the boson--exchange or phenomenological potentials,
in which the parameters are tuned in a way that no such additional
bound states appear.  There is also a
more microscopic argument. In a model allowing for two--boson
exchange (like the Bonn model discussed in refs.~\cite{HME})
such strongly attractive contributions stem from TPE (with intermediate
deltas) which are 
in low angular momentum partial waves
almost completely cancelled by two--boson--exchange graphs,
in which one of the pions is substituted by a $\rho$--meson, see
e.g.~\cite{HME} and fig.~\ref{fig:1}. It is even stated
in that last reference that ``the $2\pi$--contribution appears, in
general, too attractive and a consistent and quantitative description
of all phase shifts can never be reached''.
Further  work on a detailed understanding of correlated $\pi\rho$ exchange has
been performed by Holinde and collaborators, see~\cite{Juelpirho}.
In the EFT approach, the precise order in which the $\pi \rho$ diagrams
with intermediate deltas start contributing to
four--nucleon operators depends on the representation of the vector
fields (and thus need not appear at the same order as the
corresponding $2\pi$ graphs). Therefore, we have
constructed new chiral potentials at NNLO where in the NNLO TPE graphs
we have substituted the $\pi N$ LECs,
\beq\label{ci}
c_i \to \tilde{c}_i = c_i - c_i^\Delta~, \quad i=3,4~,
\eeq
using the formalism of ref.~\cite{BKMasp} to calculate the
$c_i^\Delta$. More precisely, we have allowed for some
fine tuning of the $\tilde{c}_i$ within the bounds given in that
reference. By this method, the equivalent TPE graphs with intermediate
deltas are subtracted and the aforementioned cancellations are
effectively taken into account. For a typical NNLO fit, we use
$c_1 = -0.81$~GeV$^{-1}$, $c_3 = -1.15$~GeV$^{-1}$, and
$c_4 = 1.20$~GeV$^{-1}$. From that, we obtain the NNLO TPE
contribution listed in table~\ref{tab1} (for explicit analytical
expressions, see appendix~\ref{app:TPE}).
A more detailed description of this
procedure and further justification of it  will be given in a forthcoming
publication~\cite{EGMinwork}. The so determined TPE NNLO contribution
and the corresponding LECs are displayed in table~\ref{tab1}. It is
important to note that in the cases where the TPE contribution is large,
the NNLO correction is sizeably smaller than the NLO one.
The resulting values for the LECs $C_i$ and $\tilde{C}_i$ at  NNLO
are consistent with the ones found at NLO. That is an
important result. 

\medskip
\noindent
We are now in the position to confront the LECs
determined from chiral effective field theory with the highly
successful phenomenological/meson models of the nuclear force.
Before doing that, some discussion concerning the NNLO potential constructed
in~\cite{EGMII} is in order. It is a perfectly viable scenario to use
the unsubtracted values for the $c_i$ as done there since the
resulting deep bound states do not influence the physics in the
two--nucleon system. As noted already, a direct comparison of the
contact terms in the potential with the ones obtained from the
meson--exchange or phenomenological approaches can not be made. 
While the physics in three-- or four--nucleon systems does not depend
on the choice of the unsubtracted or subtracted $c_i$, the latter
choice is closer to standard nuclear physics in which three--body
forces lead to small binding energy corrections \cite{EGMinwork}. In fact,
applying directly the potential from ref.\cite{EGMII} to such systems
leads to much smaller binding energies from
the two--nucleon forces alone. However, such a separation of the total
binding energy into NN and 3N contributions 
is not observable and therefore this scenario is not ruled out.
These topics will be discussed in much more detail in
\cite{EGMinwork}. At this stage, both options discussed here are viable.
It is fair to say that more detailed calculations in few--nucleon
systems have to be performed to ultimately clarify this issue.
We proceed using the modified
dimension two pion--nucleon couplings, cf. eq.(\ref{ci}).

\section{LECs from boson--exchange and phenomenological NN potentials}
\label{sec:pot}

\begin{figure}[tb]
\centerline{\epsfig{file=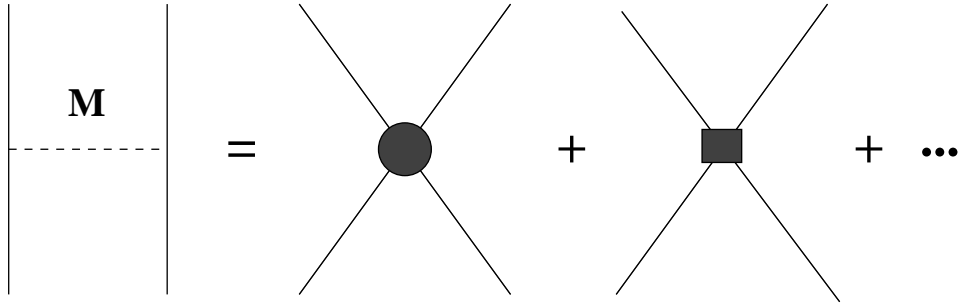,width=5in}}

\vspace{1.3cm}

\caption{\protect \small
Expansion of a meson exchange diagram in terms of local four--nucleon
operators. The dashed and solid lines denote the meson $M=\rho, \sigma,
\omega,\ldots$ and the
nucleons, respectively. The blob and the square denote insertions
with zero and two derivatives, in order. The ellipses stands for
operators with more derivatives.}\label{fig:M}
\end{figure}

\noindent
We consider first genuine one--boson--exchange models
of the NN force, in which the long range part of the interaction is given
by OPE (including in general a pion--nucleon form factor) whereas shorter
distance physics is expressed in terms of a sum over heavier mesons,
\beq
V_{\rm NN} = V_\pi + \sum_{M=\sigma, \rho, \ldots} V_M~,
\eeq
where some mesons can be linked to real resonances (like e.g. the
$\rho$--meson) or are parametrizations of certain physical effects, e.g. the
light scalar--isoscalar $\sigma$--meson is needed to supply the intermediate
range attraction (but it is not a resonance).
The corresponding meson--nucleon vertices are given
in terms of one (or two) coupling constant(s) and corresponding form factor(s),
characterized by some cut--off $\Lambda_M$. These form factors are needed to
regularize the potential at small distances (large momenta) but they
should not be given a physical interpretation. As depicted in Fig.\ref{fig:M},
in the
limit of large meson masses, keeping the ratio of coupling constant to mass
fixed, one can interpret such exchange diagrams as a sum of local operators
with increasing number of derivatives (momentum insertions).
\noindent In a highly symbolic relativistic notation, this reads,
\beq\label{reso}
(\bar N P_i N) \left( {g^2 \, \delta^{ij}\over M_R^2-t} \right)
(\bar N P_j N)
= \left( {g^2 \over M_R^2} \right) (\bar N P_i N)  (\bar N P^i N)
+ \left( {g^2\, t \over M_R^4} \right) (\bar N P_i N)
(\bar N P^i N) + \dots~,
\eeq
where the $P_i$ are projectors on the appropriate
quantum numbers for a given meson exchange (including also Dirac matrices if
needed)  and $M_R$ is the mass of corresponding heavy meson.
It should be kept in mind here that one usually makes use of
the non--relativistic expansion, i.e. the Dirac spinors on the right--hand
side of eq.~(\ref{reso}) coincide with the Pauli spinors.
In case of a momentum--dependent meson--nucleon coupling, like e.g. for
a monopole form factor normalized to one at $t=0$,
\beq
g(t) = g \, {\Lambda_M^2 \over \Lambda_M^2 - t}~,
\eeq
then the coefficient of the first $t$--dependent term in Eq.(\ref{reso})
is modified to
\beq
{g^2 \over M_R^4} \to {g^2 \over M_R^2}\, \left( {1\over M_R^2} + {2 \over
\Lambda_M^2} \right) ~,
\eeq
and accordingly for other types of form factors (dipole, monopole normalized
to one at $t=M_R^2$, etc.). The coupling constants are either determined in the
fit to the NN scattering  and bound state data or are taken from other
sources, the form factor cut--offs always have to be determined
from the fit. It is obvious from these considerations that such heavy meson
exchanges
generate four--nucleon terms with zero, two, four, $\ldots$ derivatives.
In  appendix~\ref{app:OBE}, we collect the explicit formulae for scalar,
pseudoscalar and
vector meson exchanges, which can be applied to any of the OBE potentials
by using the appropriate masses and coupling constants (and should be used
instead of the symbolic formulae given before).
As a typical example for an OBE potential we consider the Bonn-B variant
\cite{mach}. Its short range part is build from scalar ($\sigma$, $\delta$),
pseudoscalar ($\eta$) and vector meson ($\rho$, $\omega$) exchanges, and the
pertinent contributions to the LECs $C_i, \, \tilde{C}_i$ are listed in
table~\ref{tab2}. Another (more recent) OBE potential is the Nijmegen 93
potential
(denoted Nijm--93)~\cite{Nijm93}.
The Nijmegen 93 potential is particular since it also includes  mesons
with strange quarks but total strangeness zero
(like the scalar $\epsilon (1300)$ or the $\phi (1020)$) and a low--energy
representation of the Pomeron, which usually is needed to describe very high
energetic proton--proton scattering. SU(3) flavor symmetry is imposed so that
certain couplings are linked. The various contributions to the LECs
are displayed in table~\ref{tab3}.  Some of the individual terms
are unnaturally large (in particular the ones from the Pomeron), but the total
contribution of the scalar  sector is quite similar to the ones in
the Bonn-B potential, as comparison of tables~\ref{tab2},\ref{tab3} reveals.
The vector meson contributions ($\rho, \omega$) are very similar for both
potentials.
The resulting LECs for  these two OBE
potentials are summarized in table~\ref{tab4}. While there is some spread in
some
of the partial waves, the overall agreement with the LECs determined using
the chiral EFT potential listed in table~\ref{tab1} is rather satisfactory but
these results are also somewhat surprising, because the
phenomenological potential models are not constructed based on any
power counting nor chiral symmetry, plus in many cases contain
quantum field theoretically ill--defined form factors.
Still, it is gratifying to see that the contact part of the NN
potential does not depend on how the
short distance physics is parametrized.
To our knowledge, this is the first time that a
direct link between the Weinberg program of systematically deriving nuclear
forces from
chiral Lagrangians to these phenomenologically successful potentials has been
achieved in a truely quantitative manner.

\begin{figure}[tb]
\centerline{\epsfig{file=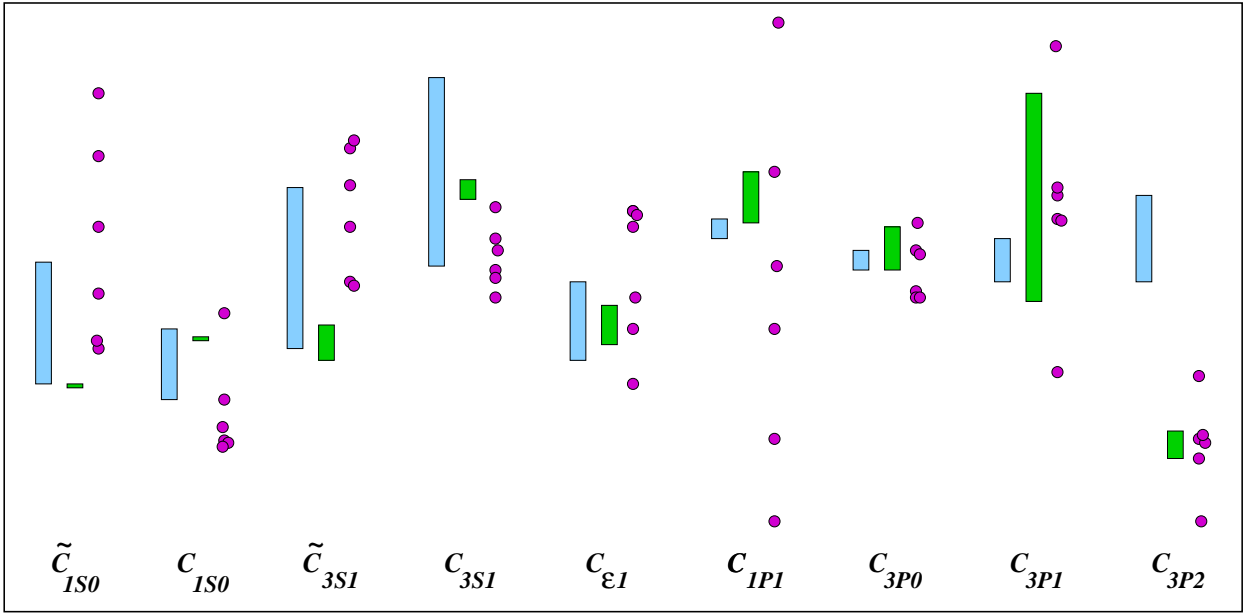,width=6.5in}}

\vspace{1.5cm}

\caption{\protect \small
Graphical representation to compare the LECs determined in EFT with
the values obtained from the various models. The left bar refers to
NLO, the middle bar to NNLO and the filled circles to the various
potentials discussed in the text. Note that the units are arbitrary,
i.e. different scale factors (including in some cases an overall minus
sign) have been assigned to the various LECs to obtain a more uniform
representation.
}\label{fig:L}
\end{figure}

\medskip
\noindent There exists also a different class of potentials, which are
constructed to give $\chi^2/{\rm datum} \simeq 1$ fits to the NN data base like
the high--precision charge--dependent CD--Bonn~2000~\cite{CDBonn}.
It contains two scalar--isoscalar
mesons in each partial wave up to angular momentum $J=5$ with the mass and
coupling constant of the second $\sigma$ fine tuned in any partial wave.
The other high--precision potentials are
the Nijmegen~I,II~\cite{Nijm93} as well as the
Argonne V18 (AV--18)~\cite{AV18}  potentials. For the former,
one--pion exchange is supplemented by heavy boson exchanges with
adjustable parameters which are fitted for all (low) partial waves
separately. The AV--18 potential starts from a very general operator
structure in coordinate space and has fit functions for all these
various operators. Note that we have switched off the various
electromagnetic corrections implemented in the  AV--18 potential code.
Such type of potentials can also be expanded in terms of
four--nucleon contact operators with increasing dimension.
We do not give the details here but only mention that we have done
this using numerical methods. The resulting LECs $C_i$ and
$\tilde{C}_i$ are also listed in table~\ref{tab4}. In fig.\ref{fig:L} we give
a graphical representation of the LECs obtained in EFT, cf. table~\ref{tab1},
compared
with the results from the six potential models considered here, see
table~\ref{tab4}.
Note that the various LECs are scaled with different factors, sometimes
including
an overall minus sign. This is done to achieve a more uniform representation.
Note also
that the theoretical uncertainties for the LECs
determined in EFT are a) small compared to their
average values and b) are smaller than the band spanned by the potential models
(even
if one only includes the high--precision ones).

\section{Naturalness of the LECs and Wigner Symmetry}
\label{sec:nat}
\noindent
First, we wish to investigate whether the LECs determined in
section~\ref{sec:eft} are of natural size. In the present context of
Weinberg power counting, dimensional scaling arguments allow one to express
any term of the effective Lagrangian with nucleon and pion fields
as well as derivative and pion mass insertions (for a derivation and
further discussion, see e.g.~\cite{F}) as
\beq
{\cal L} = c_{lmn} \, \left( {N^\dagger (\ldots ) N \over f_\pi^2 \,
  \Lambda_\chi} \right)^l \, \left( {\pi \over f_\pi} \right)^m \,
\left( {\partial^\mu, M_\pi \over \Lambda_\chi }\right)^n \, f_\pi^2
\,  \Lambda_\chi^2 ~~,
\eeq
where the $c_{lmn}$ are dimensionless numbers and $l,m,n$ are
non--negative integers. Here, $2l$ counts the number of nucleon
fields, $m$ the number of pions and $n$ the number of derivatives
or pion mass insertions. All nucleon isospin operators
and so on are non--essential to this formula and indicated by the
ellipsis. Note that this naive power counting can not be applied
to cases with spurious bound states, as witnessed by the so--called
limit cycle behaviour~\cite{BHvK,Bint,EGMcycle}. Here, we only
consider potentials with no such spurious bound states, thus the relevant scale
for
the four--nucleon interactions without derivatives ($l=2,m=n=0$)
is the inverse of the pion decay constant, $f_\pi
= 92.4\,$MeV, squared and two derivative terms ($l=2,m=0, n=2$)
are suppressed by two inverse powers of
the chiral scale $\Lambda_\chi \simeq 1$~GeV. For the LECs from the
Lagrangian (\ref{Leff}) naturalness thus amounts to
\beq
C_i \sim {c_{200} \over f_\pi^2}~, \quad (i = S, T)~, \qquad
C_j \sim {c_{202} \over f_\pi^2 \, \Lambda_\chi^2}~, \quad (i = 1, \dots, 7)~,
\eeq
and the $c_{200}$ and $c_{202}$ should be numbers of order one (if there is not
some suppression due to some symmetry, see below). Such arguments can,
of course, not say anything about the signs of the LECs. Also, it is
important to realize the prefactors which accompany the various terms
of the Lagrangian. E.g. there is
a relative factor of four in the momentum space representation between
terms $\sim {\vec q}\,^2 = (\vec{p}\,' -\vec{p}\,)^2$ and
$\sim \vec{k}\,^2 = (\vec{p}\,' +\vec{p}\,)^2/4$. Such
factors need to be accounted for. Consequently, we give in
table~\ref{tabN} the corresponding coefficients $c_{200}$ and $c_{202}$ of
the LECs as deduced from our NLO and NNLO fits using eqs.(\ref{VC}).
Inspection of the table reveals that the numbers fluctuate
between 0.3 and 3.5, i.e. the values found for these LECs are
indeed natural, with the notable exception of $f_\pi^2 \, C_T$,
which is much smaller than one (except for the upper limit of NLO
cut--offs, which is already close to the edge of having stable
fits, see also the discusion in~\cite{EGMII}).
As just mentioned, symmetry can lead to the
suppression (or enhancement) of certain coupling constants. In fact,
65 years ago Wigner~\cite{Wigrot} proposed that SU(4) spin--isospin
transformations are an approximate symmetry of the strong interactions. Such
a transformation has the form
\beq\label{Wigner}
\delta \, N = i \varepsilon_{\mu\nu} \, \sigma^\mu \, \tau^\nu \, N~,
\qquad N=\bigg(\begin{array}{c} p \\ n \end{array} \bigg) ~,
\qquad \mu, \nu =0,1,2,3~,
\eeq
with $\sigma^\mu = (1,\vec\sigma)$, $\tau^\nu=(1,\vec\tau)$, and
$\varepsilon_{\mu\nu}$ are infinitesimal group parameters. This
symmetry emerges in the large number of color limits of QCD~\cite{NC} and thus
features prominently in the nuclear forces derived from Skyrme type
models. It was recently shown~\cite{MSW} that in the limit where the S--wave
scattering lengths $a_{1S0}$ and $a_{3S1}$ go to infinity, the leading
terms in the EFT for strong NN interactions (with pions treated
perturbatively) are invariant under Wigner's SU(4) spin--isospin
transformations. This can be seen most easily from the leading
four--nucleon operators as used here, see eq.(\ref{Leff})
In this basis, the first term is clearly invariant under Wigner
transformations, cf eq.(\ref{Wigner}), whereas the second term
$\sim C_T$ obviously breaks the SU(4) symmetry. In the Weinberg
approach employed here, the leading order potential consists of these
two  four--nucleon operators supplemented by the one--pion exchange.
Still, the Wigner symmetry is kept intact to a good precision since the
resulting
fit values for $C_T$ are sizeably smaller than the corresponding ones
for $C_S$, see table~\ref{tabN}. Stated differently, $C_T$ is
unnaturally small because of the Wigner symmetry. This can be
understood from the fact that at very low energies, where one is
essentially sensitive to the (S--wave) scattering lengths, the pion--exchange
contribution can be expanded in powers of momenta, leading to terms
with at least two derivatives, see appendix~\ref{app:OBE}. One thus
effectively recovers the situation eluded to in ref.\cite{MSW}.
However, for larger momenta (say of the order of the pion mass),
the non--perturbative treatment of the pions as proposed by Weinberg
is mandatory.

\section{Conclusions and Summary}
\label{sec:summ}

\noindent
In this manuscript, we have investigated the low--energy constants
with zero and two derivatives that appear in the four--nucleon contact
interactions of the chiral effective Lagrangian for the
nucleon--nucleon forces. Our main findings can be summarized as
follows:
\begin{itemize}
\item[(1)] We have determined the LECs for the NLO and NNLO
  potentials, including the dominant charge--dependence effect
  from the pion mass difference in the one--pion exchange. To avoid
  the unphysical bound states at NNLO, we have argued that one has
  to subtract the delta contribution from the dimension two
  pion--nucleon LECs. This is in agreement with two--boson--exchange
  models, where the two--pion--exchange contribution is cancelled
  largely by $\pi \rho$ graphs.
\item[(2)] We have shown how to deduce similar type of contact
  operators from boson--exchange models in the limit of large meson
  masses. This allows to calculate the LECs in terms of meson--nucleon
 coupling constants, meson masses and (unobservable) cut--off masses.
 In a similar manner, one can examine the so--called high--precision
 potential models. We have found that in all cases, the LECs
  determined from these models are close to the values found in EFT,
  which can be considered as a new form of resonance saturation.
\item[(3)] We have shown that with the exception of one dimension
 zero coupling (the LEC $C_T$), all LECs are of natural size. The
 smallness of $C_T$ is due to Wigner's spin--isospin symmetry, as
 was already pointed out for the case of a theory with pions
 integrated out or treated perturbatively.
\end{itemize}

\noindent Clearly, these findings have further--reaching consequences.
On one side, they might allow to further constrain models of the
nucleon--nucleon interaction applicable at energies where the EFT description
can
not be used. On the other hand, in case of external sources (like e.g.
photons) or multi--nucleon operators (as they appear e.g. in the
description of the three--body forces), these considerations will
allow to at least estimate novel LECs that will appear.
In the latter case of three-- and more--nucleon systems
performing a direct fit with 5 adjustable parameters (if the
leading non--vanishing three--nucleon force is included) to 3N observables
will be a very expensive task with respect to computer power. Therefore it
might be
very helpful to have a rough estimation for the values of various couplings
appearing in
the 3N force.

\acknowledgments
\noindent
We would like to thank Hiroyuki Kamada and Andreas Nogga for many helpful
discussions
and for providing us with computer codes of the various potentials.

\appendix
\section{Reduction of the two--pion exchange contributions}\label{app:TPE}
\noindent
As stated before, we have to add the contribution of the TPE to the LECs
so as to be able to compare with the boson exchange potentials. The explicit
expressions for the renormalized TPE potential at NLO can be found in
ref.\cite{EGMI}.
Expanding those in powers of $\vec{q}$ and $\vec{k}$ allows for a
mapping on the spectroscopic LECs (of course, the TPE contains many other
contributions, which are, however, of no relevance for this discussion).
We get
\beqa
\tilde{C}_{1S0}^{\rm NLO} &=&
-\frac{1}{3}\tilde{C}_{3S1}^{\rm NLO}
= \frac{(1 + 4  g_A^2 - 8 g_A^4) M_\pi^2}{24 \pi f_\pi^4}~, \no\\
C_{1S0}^{\rm NLO} &=&
 \frac{2 + 17 g_A^2 - 88 g_A^4 }{144 \pi f_\pi^4}~, \no\\
C_{3S1}^{\rm NLO} &=&
- \frac{2 + 17 g_A^2 - 40 g_A^4 }{48 \pi f_\pi^4}~, \no\\
{C}_{\epsilon 1}^{\rm NLO} &=&
- \frac{g_A^4 }{4\sqrt{2} \pi f_\pi^4}~, \no\\
C_{1P1}^{\rm NLO} &=&
\frac{2 + 17 g_A^2 - 16 g_A^4 }{72 \pi f_\pi^4}~, \no\\
C_{3P0}^{\rm NLO} &=&
-\frac{2 + 17 g_A^2 + 74 g_A^4 }{216 \pi f_\pi^4}~, \no\\
C_{3P1}^{\rm NLO} &=&
-\frac{2 + 17 g_A^2 - 61 g_A^4 }{216 \pi f_\pi^4}~, \no\\
C_{3P2}^{\rm NLO} &=&
-\frac{2 + 17 g_A^2 - 7 g_A^4 }{216 \pi f_\pi^4}~~.
\eeqa
Note that in the chiral limit, the two leading contact interactions
do not get renormalized by TPE. Furthermore, these expressions only
depend on the lowest order pion--nucleon coupling $\sim g_A$ (or, by
virtue of the Goldberger--Treiman relation, on $g_{\pi NN}$). Similarly,
we can give the additional TPE NNLO contributions to the various LECs (for an
explicit expression of the renormalized NNLO TPE potential,
see e.g. ref.\cite{EGMII}),
\beqa
\tilde{C}_{1S0}^{\rm NNLO} &=&
\frac{g_A^2(-16 + 192m(-2c_1+ c_3) + 25 g_A^2) M_\pi^3}{256 m f_\pi^4}~, \no\\
\tilde{C}_{3S1}^{\rm NNLO} &=&
\frac{3g_A^2(16 + 64 m(-2c_1+ c_3) - 21 g_A^2) M_\pi^3}{256 m f_\pi^4}~, \no\\
C_{1S0}^{\rm NNLO} &=&
\frac{g_A^2(-368 -192 m(10 c_1 - 11 c_3 + 4c_4) + 869 g_A^2) M_\pi}
{3072 m f_\pi^4}~, \no\\
C_{3S1}^{\rm NNLO} &=&
-\frac{g_A^2(16(-7 + 4 m (10 c_1 - 11 c_3 + 4c_4)) + 81 g_A^2) M_\pi}
{1024 m f_\pi^4}~, \no\\
{C}_{\epsilon 1}^{\rm NNLO} &=&
\frac{g_A^2(8 + 32 m c_4 - 7 g_A^2) M_\pi}{64 \sqrt{2} m f_\pi^4}~, \no\\
C_{1P1}^{\rm NNLO} &=&
\frac{g_A^2(-368 + 64 m (10 c_1 - 11 c_3 - 12c_4) + 305 g_A^2) M_\pi}
{1536 m f_\pi^4}~, \no\\
C_{3P0}^{\rm NNLO} &=&
\frac{g_A^2(176 + 192 m(10 c_1 - 11 c_3 - 8 c_4) + 691 g_A^2) M_\pi}
{4608 m f_\pi^4}~, \no\\
C_{3P1}^{\rm NNLO} &=&
\frac{g_A^2(464 + 192 m(10 c_1 - 11 c_3 + 2 c_4) - 545 g_A^2) M_\pi}
{4608 m f_\pi^4}~, \no\\
C_{3P2}^{\rm NNLO} &=&
-\frac{g_A^2(112 + 192 m( -10 c_1 + 11 c_3 + 2 c_4) + 281 g_A^2) M_\pi}
{4608 m f_\pi^4}~, \no\\
\eeqa
with $m$ the nucleon mass.
These expressions depend on the dimension two LECs $c_{1,3,4}$ as discussed
before. We note that all these contributions vanish in the chiral limit.

\section{Reduction of one--boson--exchanges}\label{app:OBE}
\noindent
Here, we give the explicit expression for scalar, pseudoscalar and
vector meson exchange contributions to four--nucleon operators
with zero or two derivatives, as depicted in fig.~\ref{fig:M}.
Note that we will also include $1/m$ as well as $1/m^2$ corrections,
which are, strictly speaking, of higher orders in the power counting scheme we
are
working with and not (or partly) present in the NLO and NNLO potentials.
There is, however, no contradiction since
adding or subtracting those terms from the potential would lead to
changes smaller than the level of accuracy of our approach.
The contributions for a particular OBE potential can be obtained
by using the appropriate masses, coupling constants (and form factors)
employed there. To obtain the most general expressions, we include any
form factor as
\beq
F_M (\vec{q}\, ^2) = \alpha_1 + \alpha_2 \frac{\vec{q} \, ^2}{\Lambda_M^2} + {\cal O}
(\vec{q} \, ^4)~,
\eeq
where the coefficient $\alpha_1 =1 $ if the form factor is normalized
to one at $\vec{q} \, ^2 =0$ or $\alpha_1 \ne 1$ if the form factor is normalized
to one at $\vec{q} \, ^2 = - M^2_M$ (with $M_M$ the mass of the meson under
consideration) or it might include the meson--nucleon coupling constant
$g_M$. We will give generic expression where the corresponding
vertices are written as $g_M \times F_M (\vec{q} \, ^2)$, with $F_M$ expanded as
just discussed. Note for the  exchange of isovector bosons like the
$\pi$ or the $\rho$, the given expressions have to be multiplied by a
factor $\fet{\tau}_1 \cdot \fet{\tau}_2$ leading to a factor of (-3)
for the $T = 0$ potential considered here.

\subsection{Scalar meson exchange}
\noindent
The Lagrangian for coupling of an scalar--isoscalar meson with mass
$M_S$ and coupling constant $g_S$ reads
\beq
{\cal L}_S = g_S \bar \psi \psi \phi~,
\eeq
where $\psi$ denotes the relativistic nucleon field and $\phi$
the scalar meson (for an isovector, one simply replaces $\phi$
by $\fet \tau \cdot \fet{\phi}$, with $\tau^k$ ($k=1,2,3$) the
usual Pauli isospin matrices). In the non--relativistic expansion,
the momentum space expression for the corresponding
exchange potential with a  form factor (if applicable) characterized
by the  cut--off $\Lambda_S$ reads up to terms of order $1/m^2$
\beq
V_S (\vec{q}\, , \vec{k}\,)  = -\frac{g_S^2}{\vec{q}\,^2 + M_S^2} \,
\left[ 1 - \frac{\vec{k}\,^2}{2m^2} + \frac{\vec{q}\,^2}{8m^2} -
\frac{i}{2m^2} \vec{S} \cdot ( \vec{q} \times \vec{k} \,)  \right] \,
F_S^2  (\vec{q}\,^2)~,
\eeq
where $\vec{S} = (\vec{\sigma}_1 + \vec{\sigma}_2 )/2$ is the total
spin of the two--nucleon system. The fully relativistic form of this
exchange can be found e.g. in ref.\cite{mach}. This gives the
following contributions to the spectroscopic LECs:
\beqa
\tilde{C}_{1S0}^S &=& \tilde{C}_{3S1}^S = - \frac{4 \pi g_S^2
\alpha_1^2}{M_S^2}~, \no\\
C_{1S0}^S &=& C_{3S1}^S = \frac{4 \pi g_S^2 \alpha_1 \, (-2M_S^2 \alpha_2 +
  \alpha_1  \Lambda_S^2)}{M_S^4 \, \Lambda_S^2}~, \no\\
{C}_{\epsilon 1}^S &=&  0~, \no\\
C_{1P1}^S &=&  \frac{2 \pi g_S^2 \alpha_1 \, ( (-4+M_S^2/m^2) \alpha_1 +
8M_S^2  \alpha_2  / \Lambda_S^2)}{3 M_S^4}~, \no\\
C_{3P0}^S &=& \frac{2 \pi g_S^2 \alpha_1 \, (8 m^2 M_S^2 \alpha_2 -
(4m^2 - 3M_S^2)  \alpha_1  \Lambda_S^2)}{3 m^2 M_S^4 \, \Lambda_S^2}~, \no\\
C_{3P1}^S &=&  \frac{4 \pi g_S^2 \alpha_1 \, ( (-2+M_S^2/m^2) \alpha_1 +
4M_S^2  \alpha_2  / \Lambda_S^2)}{3 M_S^4}~, \no\\
C_{3P2}^S &=&  \frac{8 \pi g_S^2 \alpha_1 \, (2M_S^2 \alpha_2 -
  \alpha_1  \Lambda_S^2)}{3 M_S^4 \, \Lambda_S^2}~.
\eeqa

\subsection{Pseudoscalar meson exchange}
\noindent
The Lagrangian for coupling of a scalar--pseudoscalar meson with mass
$M_P$ and coupling constant $g_P$ reads
\beq
{\cal L}_P = -g_P \bar \psi i \gamma^5 \psi \pi~,
\eeq
where  $\pi$ denotes the pseudoscalar meson (for an isovector,
one simply replaces $\pi$ by $\fet \tau \cdot \fet \pi$). This
is the so--called pseudoscalar coupling.
Equivalently, one can also use a derivative type (pseudovector) coupling
\beq
{\cal L}_P' = -\frac{f_P}{M_P} \bar \psi \gamma^5 \gamma^\mu \psi
\partial_\mu \pi~.
\eeq
At tree level, these couplings are equivalent provided $g_P / m =
f_P/M_P$. Of course, chiral symmetry enforces the derivative
coupling for the Goldstone bosons. In the non--relativistic expansion,
the momentum space expression for the corresponding
exchange potential 
%(using the pseudoscalar coupling)
with a  form factor (if applicable) characterized
by the  cut--off $\Lambda_P$
reads up to terms of order $1/m^2$
\beq
V_P (\vec{q}\,)  = -\frac{g_P^2}{4m^2} \,
\frac{(\vec{\sigma}_1 \cdot \vec{q} \,)(\vec{\sigma}_2 \cdot \vec{q}
  \,)}{\vec{q}\,^2 + M_P^2}\, F_P^2 (\vec{q}\,^2)~.
\eeq
Again, the fully relativistic form of this
exchange can be found e.g. in ref.\cite{mach}. This gives the
following contributions to the spectroscopic LECs:
\beqa
\tilde{C}_{1S0}^P &=& \tilde{C}_{3S1}^P = 0~,\no \\
C_{1S0}^P &=& -3 C_{3S1}^P = \Gamma_P~, \no \\
-\frac{3}{2\sqrt{2}} C_{\epsilon 1}^P &=& -\frac{3}{2}C_{1P1}^P
= -\frac{1}{2}C_{3P0}^P = \frac{3}{4} C_{3P1}^P = \Gamma_P ~,\no\\
C_{3P2}^P &=& 0~,
\eeqa
with
\beq
\Gamma_P =  \frac{ \pi g_P^2 \alpha_1^2}{m^2 M_P^2}~.
\eeq

\subsection{Vector meson exchange}
\noindent
The Lagrangian for coupling of a vector meson with mass
$M_V$ and coupling constants $g_V$ (vector coupling) and
$f_V$ (tensor coupling) reads
\beq
{\cal L}_V = -g_V \bar \psi  \gamma^\mu \psi \phi_\mu
- \frac{f_V}{4m} \bar \psi  \sigma^{\mu\nu} \psi (\partial_\mu
\phi_\nu - \partial_\nu \phi_\mu)~,
\eeq
where  $\phi_\mu$ denotes the isoscalar--vector meson (for an isovector,
one simply replaces $\phi_\mu$ by $\fet \tau \cdot \fet \phi_\mu$).
In the non--relativistic expansion,
the momentum space expression for the corresponding
exchange potential  with a form factor (if applicable) characterized
by the  cut--off $\Lambda_V$ reads up to terms of order $1/m^2$
\beqa
V_V (\vec{q}\, , \vec{k}\,)  &=& \frac{1}{\vec{q}\,^2 + M_V^2} \,
\biggl\{ g_V^2
\biggl[ 1 + \frac{3 \vec{k}\,^2}{2m^2} - \frac{\vec{q}\,^2}{8m^2} +
\frac{3i}{2m^2} \vec{S} \cdot ( \vec{q} \times \vec{k} \,)  -
\vec{\sigma}_1 \cdot \vec{\sigma}_2 \frac{\vec{q}\,^2}{4m^2} +
\frac{1}{4m^2} (\vec{\sigma}_1 \cdot \vec{q} \,)(\vec{\sigma}_2 \cdot \vec{q}
  \,) \biggr] \no \\
&& \qquad\qquad + \frac{g_V f_V}{2m} \biggl[ - \frac{\vec{q}\,^2}{m} +
\frac{4i}{m} \vec{S} \cdot ( \vec{q} \times \vec{k} \,)
- \vec{\sigma}_1 \cdot \vec{\sigma}_2 \frac{\vec{q}\,^2}{m} +
\frac{1}{m}  (\vec{\sigma}_1 \cdot \vec{q} \,)(\vec{\sigma}_2 \cdot \vec{q}
  \,) \biggr] \no \\
&& \qquad\qquad + \frac{f_V^2}{4m^2} \left[ - \vec{\sigma}_1 \cdot
\vec{\sigma}_2
\vec{q}\,^2 + (\vec{\sigma}_1 \cdot \vec{q} \,)(\vec{\sigma}_2 \cdot \vec{q}
\,)
\right]
\biggr\}~F_V^2 (\vec{q}\,^2)~,
\eeqa
Again, the fully relativistic form of this
exchange can be found e.g. in ref.\cite{mach}. This gives the
following contributions to the spectroscopic LECs:
\beqa
\tilde{C}_{1S0}^V &=& \tilde{C}_{3S1}^V = \frac{4 \pi g_V^2
\alpha_1^2}{M_V^2}~, \no\\
C_{1S0}^V &=&  \frac{ \pi  \alpha_1 \, (8 m^2 g_V^2 M_V^2 \alpha_2 -
(4 m^2 g_V^2 -  (2f_V^2 + 2f_V g_V + 3g_V^2) M_V^2 )
  \alpha_1  \Lambda_V^2)}{m^2 M_V^4 \, \Lambda_V^2}~, \no\\
C_{3S1}^V &=&  \frac{ \pi  \alpha_1 \, (24 m^2 g_V^2 M_V^2 \alpha_2 -
(12 m^2 g_V^2 +  (2f_V^2 + 10 f_V g_V - g_V^2) M_V^2 )
  \alpha_1  \Lambda_V^2)}{3m^2 M_V^4 \, \Lambda_V^2}~, \no\\
{C}_{\epsilon 1}^V &=&  \frac{2 \sqrt{2} \pi (f_V+g_V)^2 \alpha_1^2}{3
  m^2 M_V^2}~, \no\\
C_{1P1}^V &=&  \frac{ 4\pi  \alpha_1 \, (-4 m^2 g_V^2 M_V^2 \alpha_2 +
(2 m^2 g_V^2 - f_V  (f_V + g_V ) M_V^2)
  \alpha_1  \Lambda_V^2)}{3m^2 M_V^4 \, \Lambda_V^2}~, \no\\
C_{3P0}^V &=&  \frac{ 4 \pi  \alpha_1 \, (-4 m^2 g_V^2 M_V^2 \alpha_2 +
(2 m^2 g_V^2 +  (2f_V^2 + 9 f_V g_V +6 g_V^2) M_V^2 )
  \alpha_1  \Lambda_V^2)}{3m^2 M_V^4 \, \Lambda_V^2}~, \no\\
C_{3P1}^V &=&  \frac{ 2 \pi  \alpha_1 \, (-8 m^2 g_V^2 M_V^2 \alpha_2 +
(- M_V^2 f_V^2 +  4g_V (m^2 g_V + ( f_V+ g_V) M_V^2) )
  \alpha_1  \Lambda_V^2)}{3m^2 M_V^4 \, \Lambda_V^2}~, \no\\
C_{3P2}^V &=&  \frac{2 \pi  \alpha_1 \, (f_V^2 M_V^2 \alpha_1 / m^2 +
 4g_V^2 ( \alpha_1 - 2M_V^2 \alpha_2/ \Lambda_V^2))}{3 M_V^4}~.
\eeqa

\vfill\eject

%%%%%%%%%%%%%%%%%%%%%%%%%%%%%%% refs %%%%%%%%%%%%%%%%%%%%%%%%%%%%%%%%%%%%%

\pagebreak

\centerline {{\large \bf TABLES}}

\vspace{1cm}

\begin{table}[t]
\begin{tabular}{|l|cc|cc|}
LEC  &  TPE(NLO)  & TPE(NNLO) & $C_i$~(NLO) & $C_i$~(NNLO)  \\
    \hline
$\tilde{C}_{1S0}$  &  $-0.004$  &  $0.003$
             &  $-0.156 \ldots -0.110$ & $-0.160 \ldots -0.158$ \\
${C}_{1S0}$        &     $-0.585$  &  $-0.070$  &  $1.048 \ldots 1.253$ &
    $1.135 \ldots 1.134$ \\
$\tilde{C}_{3S1}$  &  $0.013$  &  $0.001$
             &  $-0.155 \ldots -0.023$ & $-0.159 \ldots -0.134$ \\
${C}_{3S1}$        &     $0.653$  &  $-0.181$  &  $0.250 \ldots 0.840$ &
    $0.637 \ldots 0.587$ \\
${C}_{\epsilon 1}$    &  $-0.195$  &  $0.117$  &  $-0.302 \ldots -0.384$ &
    $-0.369 \ldots -0.326$ \\
${C}_{1P1}$        &     $-0.069$  &  $-0.099$  &  $0.260 \ldots 0.273$ &
    $0.234 \ldots 0.268$ \\
${C}_{3P0}$        &     $-0.436$  &  $-0.071$  &  $0.800 \ldots 0.855$ &
    $0.727 \ldots 0.857$ \\
${C}_{3P1}$        &     $0.252$  &  $0.011$  &  $-0.126 \ldots -0.093$ &
    $-0.141 \ldots 0.026$ \\
${C}_{3P2}$        &     $-0.023$  &  $0.036$  &  $-0.325 \ldots -0.259$ &
    $-0.464 \ldots -0.445$ \\
  \end{tabular}
\vspace{0.3cm}
\caption{Values of the LECs at NLO and NNLO for the cut--off values
  $\Lambda =500 \ldots 600\,$MeV. Also given are the contributions from
  two--pion exchange at NLO and NNLO which are contained in the values
  of the LECs as
  explained in the text. The $\tilde{C}_i$ are in $10^{4}~$GeV$^{-2}$
  and the $C_i$ in $10^{4}~$GeV$^{-4}$.}
\label{tab1}
\end{table}

\vspace{3cm}

\begin{table}[htb]
\begin{center}
\begin{tabular}{|l||c|c|c|c|c||c|}
LEC  &  $\eta$  & $\sigma$ & $\delta$ & $\omega$ & $\rho$ & sum  \\
    \hline
$\tilde{C}_{1S0}$  &  $0.000$  &  $-0.392$
             &  $-0.023$ & $0.287$  & $0.011$ & $-0.117$ \\
${C}_{1S0}$        &  $0.033$  &  $1.513$
             &  $0.036$ & $-0.560$  & $0.254$ & $1.276$ \\
$\tilde{C}_{3S1}$  & $0.000$  &  $-0.424$
             &  $0.070$ & $0.287$  & $-0.034$ & $-0.101$ \\
${C}_{3S1}$        &  $-0.011$  &  $1.030$
             &  $-0.108$ & $-0.777$  & $0.526$ & $0.660$ \\
${C}_{\epsilon 1}$    &  $-0.032$  &  $0.000$
             &  $0.000$ & $0.077$  & $-0.455$ & $-0.410$ \\
${C}_{1P1}$        &  $-0.022$  &  $-0.607$
             &  $0.059$ & $0.536$  & $0.488$ & $0.454$ \\
${C}_{3P0}$        & $-0.067$  &  $-0.786$
             &  $-0.011$ & $1.187$  & $0.597$ & $0.921$ \\
${C}_{3P1}$        & $0.045$  &  $-0.860$
             &  $-0.015$ & $0.753$  & $0.003$ & $-0.075$ \\
${C}_{3P2}$        & $0.000$  &  $-1.008$
             &  $-0.024$ & $0.536$  & $0.101$ & $-0.396$ \\
  \end{tabular}
\vspace{0.3cm}
\caption{Contributions of the various boson exchanges to the LECs
for the Bonn-B potential and the corresponding sum.
The $\tilde{C}_i$ are in $10^{4}~$GeV$^{-2}$ and the $C_i$ in
$10^{4}~$GeV$^{-4}$.}
\label{tab2}
\end{center}
\end{table}

\begin{table}[htb]
\begin{center}
\begin{tabular}{|l||c|c|c|c|c|c|c|c|c|c|}
LEC  &  $\eta$ & $\eta '$ & $\rho$ & $\omega$ & $\phi$ & $a_0$ &
$\epsilon$ & $f_0$ & $a_2$ & Pom.\\
\hline
$\tilde{C}_{1S0}$
             &  $0.000$  & $0.000$ &$0.020$  & $0.237$ & $0.001$
             &  $-0.031$  & $-0.578$ &$-0.201$  & $0.001$ & $0.490$\\
${C}_{1S0}$
             &  $0.041$  & $0.013$ &$0.191$  & $-0.445$ & $-0.002$
             &  $0.134$  & $3.461$ &$0.867$  & $-0.005$ & $-2.829$\\
$\tilde{C}_{3S1}$
             &  $0.000$  & $0.000$ &$-0.055$  & $0.237$ & $0.001$
             &  $0.094$  & $-0.578$ &$-0.201$  & $-0.003$ & $0.490$\\
${C}_{3S1}$
             &  $-0.014$  & $-0.004$ &$0.550$  & $-0.700$ & $-0.003$
             &  $-0.403$  & $3.461$ &$0.867$  & $0.015$ & $-2.829$\\
${C}_{\epsilon 1}$
             &  $-0.038$  & $-0.012$ &$-0.383$  & $0.090$ & $0.000$
             &  $0.000$  & $0.000$ &$0.000$  & $0.000$ & $0.000$\\
${C}_{1P1}$
             &  $-0.027$  & $-0.009$ &$0.431$  & $0.423$ & $0.002$
             &  $0.250$  & $-2.194$ &$-0.539$  & $-0.010$ & $1.791$\\
${C}_{3P0}$
             &  $-0.082$  & $-0.026$ &$0.645$  & $1.167$ & $0.006$
             &  $-0.072$  & $-1.985$ &$-0.466$  & $0.003$ & $1.613$\\
${C}_{3P1}$
             &  $0.054$  & $0.017$ &$0.032$  & $0.660$ & $0.003$
             &  $-0.078$  & $-2.087$ &$-0.501$  & $0.003$ & $1.700$\\
${C}_{3P2}$
             &  $0.000$  & $0.000$ &$0.114$  & $0.457$ & $0.002$
             &  $-0.090$  & $-2.308$ &$-0.579$  & $0.003$ & $1.887$\\
  \end{tabular}
\vspace{0.3cm}
\caption{Contributions of the various boson exchanges to the LECs
for the Nijmegen 93 potential. Pseudoscalars: $\eta, \eta '$, vectors:
$\rho, \omega, \phi$, scalars: $a_0, \epsilon, f_0, a_2, {\rm Pomeron}$.
The $\tilde{C}_i$ are in $10^{4}~$GeV$^{-2}$ and the $C_i$ in
$10^{4}~$GeV$^{-4}$.}
\label{tab3}
\end{center}
\end{table}

\vspace{-1cm}

\begin{table}[bht]
\begin{center}
\begin{tabular}{|l||c|c|c|c|c|c|}
LEC  &  Bonn-B & CD-Bonn$^\star$ & Nijm-93
     & Nijm-I$^\star$ & Nijm-II$^\star$   & AV-18$^\star$
      \\
    \hline
$\tilde{C}_{1S0}$
             &  $-0.117$ & $-0.140$ &$-0.061$  & $-0.137$ & $-0.091$
             & $-0.037$ \\
${C}_{1S0}$
             &  $1.276$  & $1.388$ &$1.426$  & $1.391$ & $1.357$
             & $1.409$ \\
$\tilde{C}_{3S1}$
             &  $-0.101$ & $-0.103$ & $-0.014$  & $-0.058$ & $0.029$
             & $0.026$ \\
${C}_{3S1}$
             &  $0.660$  & $0.869$ & $0.940$  & $0.762$ & $0.795$
             & $0.867$ \\
${C}_{\epsilon 1}$
             &  $-0.410$ & $-0.315$ & $-0.343$  & $-0.221$ & $-0.241$
             & $-0.226$ \\
${C}_{1P1}$
             &  $0.454$  & $0.228$ &$0.119$  & $0.328$ & $0.401$
             & $0.290$ \\
${C}_{3P0}$
             &  $0.921$  & $0.956$ & $0.802$  & $0.802$ & $0.949$
             & $0.723$ \\
${C}_{3P1}$
             &  $-0.075$ & $-0.051$ & $-0.197$  & $-0.059$ & $-0.075$
             & $0.067$ \\
${C}_{3P2}$
             &  $-0.396$ & $-0.451$ & $-0.513$  & $-0.453$ & $-0.451$
             & $-0.467$ \\
  \end{tabular}
\vspace{0.3cm}
\caption{Results for the LECs for various OBE and other type of potentials
as explained in the text. The so--called high precision potentials are
marked by a star.
The $\tilde{C}_i$ are in $10^{4}~$GeV$^{-2}$ and the $C_i$ in
$10^{4}~$GeV$^{-4}$.}
\label{tab4}
\end{center}
\end{table}

\begin{table}[htb]
\begin{tabular}{|l|c|c|}
           &  NLO              & NNLO   \\
    \hline
$f_\pi^2 \,{C}_S$
             &  $-1.053 \ldots -0.303$ & $-1.079 \ldots -0.953$ \\
$f_\pi^2 \,{C}_T$
             &  $-0.002 \ldots  0.147$ & $0.002 \ldots 0.040$ \\
\hline
$f_\pi^2 \Lambda_\chi^2 \,{C}_1$
             &  $1.707 \ldots 3.162$ & $3.143 \ldots 2.665$ \\
$4 \,f_\pi^2 \, \Lambda_\chi^2 \,{C}_2$
             &  $1.348 \ldots 3.246$ & $2.029 \ldots 2.251$ \\
$f_\pi^2 \Lambda_\chi^2 \,{C}_3$
             &  $-0.047 \ldots -0.315$ & $0.403 \ldots 0.281$ \\
$4\, f_\pi^2 \Lambda_\chi^2 \,{C}_4$
             &  $-0.583 \ldots -0.933$ & $-0.364 \ldots -0.428$ \\
$2\, f_\pi^2 \Lambda_\chi^2 \,{C}_5$
             &  $2.418 \ldots 2.314$ & $2.846 \ldots 3.410$ \\
$f_\pi^2 \Lambda_\chi^2  \,{C}_6$
             &  $-0.385 \ldots -0.651$ & $-0.728 \ldots -0.668$ \\
$4\, f_\pi^2 \Lambda_\chi^2 \,{C}_7$
             &  $-1.790 \ldots -2.120$ & $-1.929 \ldots -1.681$ \\
  \end{tabular}
\vspace{0.3cm}
\caption{Naturalness coefficients  of the LECs at NLO and NNLO for the cut--off
values
  $\Lambda =500 \ldots 600\,$MeV.  The $f_\pi^2 \, {C}_{S,T}$  and the
 $f_\pi^2
  \Lambda_\chi^2 \,C_i$ are dimensionless.}
\label{tabN}
\end{table}

\bigskip

\centerline {{\large \bf FIGURES}}

\vspace{3cm}

\vspace{3.5cm}

\pagebreak

$\,$

\vspace{2.5cm}

\end{document}